\documentclass[aps,twocolumn,prc,epsfig,showpacs]{revtex4}
\usepackage{graphicx}
\usepackage{epsfig}
\begin{document}
\title{Radii in weakly-bound light halo nuclei}
\author{M. T. Yamashita}
\address
{Laborat\'orio do Acelerador Linear, Instituto de
F\'{\i}sica, Universidade de S\~{a}o Paulo, C.P. 66318, CEP
05315-970, S\~{a}o Paulo, Brazil}
\author{
Lauro Tomio}
\address
{Instituto de F\'\i sica Te\'orica, Universidade
Estadual Paulista, \\
Rua Pamplona, 145, Bela Vista, \\
01405-900, S\~{a}o Paulo, Brazil}
\author{T. Frederico}
\address
{Departamento de F\'\i sica, Instituto Tecnol\'ogico de
Aeron\'autica, Centro T\'ecnico Aeroespacial, 12228-900, S\~ao
Jos\'e dos Campos, Brazil}

\date{\today}

\begin{abstract}
A systematic study of the root-mean-square distance between the
constituents of weakly-bound nuclei consisting of two halo
neutrons and a core is performed using a renormalized zero-range
model. The radii are obtained from a universal scaling function
that depends on the mass ratio of the neutron and the core, as
well as on the nature of the subsystems, bound or virtual. Our
calculations are qualitatively consistent with recent data
for the neutron-neutron root-mean-square distance in the halo of
$^{11}$Li and $^{14}$Be nuclei.
\end{abstract}

\maketitle

\section{Introduction}

Light exotic-nuclei with one or two weakly bound neutrons in their
halo offer the opportunity to study large systems at small
nuclear density. (A review on the theoretical approaches and
characteristics of halo nuclei can be found in
ref.~\cite{zhukov93}.) The constituents of the halo have a high
probability to be found much beyond the interaction range. Then,
the concept of a short-range interaction between the
particles and its implications are useful in understanding the
few-body physics of the halo. The quantum description of such
large and weakly bound systems are universal and can be defined by
few physical scales despite the range and details of the pairwise
interaction~\cite{nielsen01}. The particular halo-nuclei, with a
neutron-neutron-core ($n-n-c$) structure, like $^6$He, $^{11}$Li, 
$^{14}$Be and $^{20}$C, are examples of weakly-bound three-body
systems~\cite{audi95}, where the above universal aspects can be
explored theoretically~\cite{amorim97}.

In weakly bound three-body systems, when the two-particle
s-wave scattering lengths have large magnitudes, it is possible
the occurrence of excited s-wave Efimov states~\cite{efimov70}.
It was suggested in ref.~\cite{fedorov94} that these states
could be present in some halo nuclei. This possibility was further
investigated and refined in ref.~\cite{amorim97}. A parametric
region was determined in which Efimov states can exist. Such
region, for a bound three-body system, was defined in a plane
given by the two possible (bound or virtual) subsystem energies.
The promising candidate to exhibit an excited Efimov state was
found to be $^{20}$C~\cite{amorim97, mazum00}.

A few-body system interacting through a short range force can be
parameterized by few physical scales~\cite{ad95}. For a zero-range
force in three-space dimensions, it is expected a new physical scale  
for each new particle added to the system, unless symmetry and/or
angular momentum forbids the particles to be near each other. The
three-body system in the state of zero total angular momentum, has
the bound or virtual subsystems energies and the ground state
three-body energy as the dominating physical scales. Any
observable can be expressed in terms of the ratios between the
physical scales, given by a scaling function, when the scattering
length goes to infinity, or the interaction range tends to zero
(scaling limit)~\cite{fred99,ad88}. In that sense the scaling
function is an useful tool to study three-body observables and
provides a zero order approximation to guide realistic
calculations with short-range interactions.

The scaling functions allow one to easily perform systematic
studies of various three-body halo-nuclei properties
with different types of two-body subsystems as, for example,
$^6$He, $^{11}$Li, $^{14}$Be and $^{20}$C~\cite{fb17}.
A classification scheme proposed for a bound three-body
system~\cite{jensen03}, can be investigated in terms of a scaling
function for the different average distances between the
constituents of a neutron-neutron-core halo nucleus.
The classification of the three-body halo system depends on the
kind of the two-body subsystems, bound or virtual.

For the case of two identical particles in the three-body system,
we have to consider four possibilities for the two-body subsystems
(see figure 1): all unbound ({\it Borromean}); all bound ({\it
All-bound}); one bound and two unbound ({\it Tango}~\cite{robi99}
configuration); and one unbound with two bound (we suggest a name
{\it Samba} for this configuration). One natural example of
halo-nuclei {\it Samba} system is the $^{20}$C nucleus, which is
composed by two-neutrons and a $^{18}$C core. The neutron and
$^{18}$C forms the weakly bound state of
$^{19}$C~\cite{audi95,naka99}.
\begin{figure}[thbp]
\centerline{\epsfig{figure=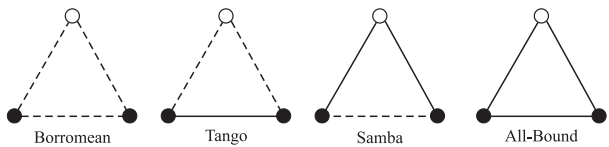,width=8cm}}
\caption[dummy0]{Diagrammatical representation of the
classification scheme for weakly-bound three-body systems.
Two-body bound state are represented with solid line, and virtual
state with dashed line.} \label{fig1}
\end{figure}

In the present work, we study the root-mean-square distances between
the constituents of a bound three-body system, where we have two
identical particles $n$ and a core named $A$. By $n$ we mean
neutron in halo nuclei, but we allow the pair neutron-neutron ($nn$) 
to be bound in order to cover all the configurations presented above.
We represent the radii in a scaling plot in terms of
a dimensionless product, extending to halo nuclei a previous
application that was done for molecules~\cite{mol03}.
Using these scaling plots we can
follow the behavior of the different radii when it happens a
transition between one kind of system to another one.
Starting from the {\it Borromean} case (all unbound), we can go to
the {\it Samba} case by increasing the binding energy of the pair
$nA$ (keeping $nn$ unbound); and to the {\it Tango} case by increasing
the binding energy of the pair $nn$ (keeping $nA$ unbound).
In particular, we calculate the mean square distance
between the two neutrons in $^6$He, $^{11}$Li and $^{14}$Be (which
have been measured recently~\cite{marq00}) from the known bound or
virtual energies of the subsystems and the two-neutron separation
energy.

The calculation of the scaling function for the different radii of
the three-body system is performed with a renormalization scheme
applied to three-body equations with $s-$wave zero-range pairwise
potential, which makes use of subtracted equations~\cite{ad95,yama02}.

As our approach is restricted to $s-$wave two-body interactions,
some limitations in the application of our analysis are expected.
The argument has the origin in the fact that, in some cases, the
two-body subsystem interaction in $p-$wave is considered to be
important for the three-body binding. The examples are
$^6$He~\cite{zheng93}, known to be bound by the $p-$wave
interaction in  $n-^4$He; and $^{11}$Li, where both $p-$ and
$s-$waves might be relevant to describe the ground state of the
unbound subsystem $^{10}$Li (see discussion and references in
\cite{caggi99}). However, as pointed out in ref.\cite{zin95} (in a
discussion related to $^{11}$Li), one should also noticed that
even the three-body wave function with $s-$wave $nn$ correlation
produces a ground state of the halo nuclei with two or more
shell-model configurations. The effect of the above $s-$wave
restriction is also reduced due to the fact that we are
considering the experimental energies in our approach, which
implicitly are carrying the effect of higher partial waves in the
interaction. Another aspect is that the radius are obtained from
the tail of the wave-function, which is dominated by the $s-$wave.

The paper is organized as follows. In section II, we present our
formalism, which contains the subtracted method to treat the
Faddeev equations with two identical particles, leading to the form
factors from which we obtain the different mean-square radii.
We also give a brief discussion of the scaling functions
to describe the radii.
In section III, we discuss the classification scheme.
In section IV, we present our numerical results for the
root-mean-square distances between the particles. Our conclusions are
summarized in section V.

\section{Formalism}

In the next subsection, we consider the formalism given in
refs.~\cite{yama02,mol03} for coupled channels to calculate the
three-body wave functions and the possible different radii with
a zero-range pairwise interaction. We solve the homogeneous
three-body Faddeev equations
for a system with two identical particles $\alpha$ and a third one
$\beta$ in a renormalized subtracted form, which allows one to obtain
the observables as a function of the two and three-body scales of
the system. The corresponding masses of the particles
$\alpha$ and $\beta$ are $m_\alpha$ and $m_\beta$.

\subsection{Subtracted Faddeev Equations}
Next, we follow the model presented in refs.~\cite{yama02,mol03} for the
subtracted Faddeev equations, and consider units such that $\hbar=
1$ and $m_\alpha=1$.  For the
subtraction energy that is required in the model, we choose
$\mu_{(3)}^ 2$.  In this case, all the energies and momentum
variables are rescaled to dimensionless variables considering this
subtraction energy. After partial wave projection, the $s-$wave
coupled subtracted integral equations for the Faddeev spectator
functions $\chi_{\alpha\alpha}$ and $\chi_{\alpha\beta}$, are
given by:
\begin{eqnarray}
\chi_{\alpha \alpha}(y)&=&2\tau_{\alpha \alpha}(y;\epsilon_3)
\int_0^\infty dx \frac{x}{y} G_1(y,x;\epsilon_3)\chi_{\alpha
\beta}(x)
\label{chi1} \\
\chi_{\alpha \beta}(y)&=&\tau_{\alpha \beta}(y;\epsilon_3)
\int_0^\infty dx \frac{x}{y} \left[G_1(x,y;\epsilon_3)
\chi_{\alpha \alpha}(x)
\right.  \nonumber \\
 &+& \left. A G_2(y,x;\epsilon_3) \chi_{\alpha \beta}(x)\right] ,
\label{chi2}
\end{eqnarray}
\begin{eqnarray}
\tau_{\alpha \alpha}(y;\epsilon_3)&\equiv &\frac{1}{\pi}
\left[\sqrt{\epsilon_3+ {\cal C}_1^A y^2} \mp
\sqrt{\epsilon_{\alpha \alpha}} \right]^{-1}, \label{tau1}
\\
\tau_{\alpha \beta}(y;\epsilon_3)&\equiv &\frac{1}{\pi}
\frac{\left( {\cal C}_2^A \right)^{3/2}} {\left[\sqrt{\epsilon_3+
{\cal C}_2^{(A+1)} y^2} \mp \sqrt{\epsilon_{\alpha \beta}}\right]}
, \label{tau2}
\\
G_1(y,x;\epsilon_3)&\equiv &\log
\frac{(\epsilon_3+x^2+xy)+{\cal C}_2^A\;y^2}
{(\epsilon_3+x^2-xy)+{\cal C}_2^A\;y^2}
\nonumber \\
&-& \log\frac{(1+x^2+xy)+{\cal C}_2^A\;y^2}
{(1+x^2-xy)+{\cal C}_2^A\;y^2} ,
\label{G1} \\
G_2(y,x;\epsilon_3)&\equiv & \log
\frac{(\epsilon_3+xy/A)+{\cal C}_2^A\;(y^2+x^2)}
     {(\epsilon_3-xy/A)+{\cal C}_2^A\;(y^2+x^2)} \nonumber
\\ &-& \log \frac{(1+xy/A)+{\cal C}_2^A\;(y^2+x^2)}
                 {(1-xy/A)+{\cal C}_2^A\;(y^2+x^2)}~,
                 \label{G2}
\end{eqnarray}
where we are defining the mass ratio and the constant mass
factors by
\begin{equation}
A\equiv \frac{m_\beta}{m_\alpha},\;\;\;\;
{\cal C}_{j=1,2}^{A}\equiv\left(\frac{j}4+\frac{1}{2A}\right)
\label{C}.
\end{equation}
The plus and minus signs in eqs.~(\ref{tau1}) and (\ref{tau2})
refer to virtual and bound two-body subsystems, respectively.
In the equations above, the dimensionless three-body energy
$\epsilon_3$ and the two-body energies
($\epsilon_{\alpha \alpha}$ and $\epsilon_{\alpha \beta}$),
are related to the corresponding physical quantities
by $\epsilon_3\equiv -E_3/\mu_{(3)}^2,$
$\epsilon_{\alpha \alpha} \equiv -E_{\alpha \alpha} /\mu^2_{(3)},$
and $\epsilon_{\alpha \beta} \equiv
-E_{\alpha \beta}/\mu^2_{(3)}.$ The three-body physical quantities
can be written in terms of a  scaling function, i.e., the
dimensionless product of the observable and some power of the
three-body binding energy $E_3$, when the value of $\mu^2_{(3)}$
is determined from the known value of $E_3$ and  consequently the
three-body quantities are naturally a function of $E_3$ and the
ratios $E_{\alpha \alpha} /E_{3}$ and $E_{\alpha\beta}/E_{3}$.
Note that we are using the same symbol $A$ for the mass ratio
as well as for the core label, considering that, in a $n-n-$core nucleus
we have the core consisting of $A$ nucleons and $A$ can also
be identified with the mass ratio ($m_\beta=A m_n$, $m_\alpha=m_n$).
However, our expression can be generally applied for non-integer
values of $A$.

For large scattering lengths, the details of the interaction are
unimportant to describe few-body systems, as the short-range
informations, beyond the two-body scattering lengths, are carried
out by one typical three-body scale. Therefore, the low-energy
observables present a scaling behavior quite universal with the
three-body binding energy~\cite{fred87,ad95}.
In the limit of infinite scattering lengths or zero range
interactions, the function which represents a given correlation
between two three-body observables, written in terms of scaled
variables, converges to a single curve, despite the existence of
many other Efimov states.

For a three-body system with binding energy $E_3$, in the
{\it scaling limit}~\cite{amorim97,yama02}, one general
three-body physical observable $\cal O$, with dimension of energy
to the power $\eta$, at a particular energy $E$, can be written
as a function $\cal F$ of the ratios between the two and three-body
energies,  such that
\begin{eqnarray}
&&{\cal{O}}\left(E, E_{3},E_{\alpha\alpha},E_{\alpha\beta},\right)
= \nonumber \\
&&(E_{3})^{\eta}
{\cal F}\left(\sqrt{\frac {E}{E_{3}}},
\pm\sqrt{\frac{E_{\alpha\alpha}}{E_{3}}},
\pm\sqrt{\frac{E_{\alpha\beta}}{E_{3}}}, A\right) \ .
\label{o}
\end{eqnarray}
The two-body energies $E_{\alpha\gamma}$ ($\gamma=\alpha,\;\beta$),
are negative quantities, corresponding to bound or virtual states.
The nature of such two-body state, bound or virtual, is revealed in the
momentum space, such that we have a bound state when $\sqrt{|E_{\alpha\beta}|}$
is positive and a virtual state when $\sqrt{|E_{\alpha\beta}|}$ is negative.
So, in equation (\ref{o}), the signs $+$ or $-$ mean a bound or virtual
two-body subsystem,
respectively. The different radii of the bound $\alpha\alpha\beta$
system are functions defined from the eq.~(\ref{o}) with $E=E_3$, which
depend on the mass ratio, $A$, the ratios of the two and
three-body energies and the kind of subsystem interactions
(bound or virtual).

\subsection{Radii calculation}
The Faddeev components of the wave-function for the
$\alpha\alpha\beta$ system are  written in terms of the spectator
functions, obtained from the solution of eqs.~(\ref{chi1}) and
(\ref{chi2}) in momentum space:
{\small
\begin{eqnarray} \nonumber
&&\Psi_{\alpha\alpha}(\vec y,\vec z)=\left(\frac{1}{\epsilon_3+
{\cal C}_{1}^{A}\; \vec y^2+\vec z^2}- \frac{1}{1+ {\cal
C}_{1}^{A}\; \vec y^2+\vec z^2}\right)
\\ \nonumber
&&\times\left[\chi_{\alpha\alpha}(|\vec y|)+
\chi_{\alpha\beta}\left(\left|\vec z-\frac{\vec
y}{2}\right|\right) + \chi_{\alpha\beta}\left(\left|\vec
z+\frac{\vec y}{2}\right|\right) \right], \\ \label{psiaa} \\
\nonumber &&\Psi_{\alpha\beta}(\vec y,\vec z)= \\ \nonumber
&&\left(\frac{1}{\epsilon_3+ {\cal C}_{2}^{A}\; \vec z^2+ {\cal
C}_{2}^{A+1}\; \vec y^2}- \frac{1}{1+ {\cal C}_{2}^{A}\; \vec z^2+
{\cal C}_{2}^{A+1}\; \vec y^2}\right) \\ \nonumber
&&\times\left[\chi_{\alpha\alpha}\left(\left|\vec z - \frac{A \vec
y}{A+1}\right|\right)+ \chi_{\alpha\beta}(|\vec y|) +
\chi_{\alpha\beta}\left(\left|\vec z+\frac{\vec
y}{A+1}\right|\right) \right], \label{psiab}
\end{eqnarray}
} where ${\cal C}_{j}^{A}$ is defined in eq.~(\ref{C}).
The Faddeev components are denoted by the sub-indices of the interacting pair.
Representing the pair by $\alpha\gamma$ with $\gamma\ = \ \alpha $
or $\beta $, one has that $\vec z$ is the relative momentum
between $\alpha$ and $\gamma$ and $\vec y$ is the relative
momentum of the third particle to the center-of-mass of the system
$\alpha \gamma$. For the halo nuclei the notation is $\alpha~=~n$
and $\beta$ is the core represented by $A$.

The momentum component of the total wave-function,
\begin{eqnarray}
\Psi_{Ann^\prime}= \Psi_{nn^\prime}+\Psi_{An}+
\Psi_{An^\prime}~,
\label{psi} \end{eqnarray}
is symmetrical by the exchange between the neutrons,
$n$ and $n^\prime$, while the antisymmetry is
given by the singlet spin-component (not explicitly shown). The
different mean-square radii are calculated from the derivative of
the Fourier transform of the respective matter density in respect
to the square of the momentum transfer. The relative mean-square
distances between the halo neutrons and between the neutron and
the core ($\gamma~=n,A$) are obtained from the expression
\begin{eqnarray}
\langle r^2_{n\gamma}\rangle= -6 \frac{dF_{n\gamma}(
q^2)}{dq^2}\bigg|_{q^2=0}, \label{rab}
\end{eqnarray}
where
\begin{eqnarray}
F_{n\gamma}( q^2)=\int\;d^3y\;d^3z
\Psi_{Ann^\prime}\left(\vec{y},\vec{z}+\frac{\vec q}{2}\right)\;
\Psi_{Ann^\prime}\left(\vec{y},\vec{z}-\frac{\vec{q}}{2}\right)
\label{F4}
\end{eqnarray}
is the Fourier transform of the two-body densities, which is
a function of the momentum transfer $\vec q$
(given in units of $\mu_{(3)}$). The relative momentum between
$n$ and $\gamma$ is $\vec{z}\pm\vec{q}/2$ and the relative
momentum of the third particle to the center-of-mass of $n\gamma$ is
$\vec{y}$.

Analogous equations can be found for the mean square distances of
the neutron and the core to the center-of-mass system
($\langle r^2\rangle_\gamma$), in terms of the Fourier transform of the one-body density.

\subsection{Scaling functions for the radii}

The scaling functions for the mean-square separation distances
between the particles in the three-body system can be written
according to eq.(\ref{o}) with $E\equiv E_3$. The scaling
functions for the radii are written as:
\begin{eqnarray}
&&\sqrt{\langle r^2_{n\gamma}\rangle |E_{3}|} ={\cal
R}_{n\gamma}\left(\pm \sqrt{\frac{E_{nn}}{E_{3}}},
\pm\sqrt{\frac{E_{nA}}{E_{3}}}, A\right) , \label{rag}
\\
&&\sqrt{\langle r^2_{\gamma}\rangle |E_{3}|}=
{\cal R}_{\gamma}^{cm}\left(\pm \sqrt{\frac{E_{nn}}{E_{3}}},
\pm\sqrt{\frac{E_{nA}}{E_{3}}}, A\right) . \label{ragcm}
\end{eqnarray}

To study the different types of three-body systems, the general
scaling function for the radii given by eqs.~(\ref{rag}) and
(\ref{ragcm}) will be studied in the configurations of figure 1.
However, one noticeable situation occurs when the Efimov limit is
reached, for which the subsystems energies vanishes, and
\begin{eqnarray}
\sqrt{\langle r^2_{n\gamma}\rangle |E_3|} ={\cal R}_{n\gamma}
\left(A\right),~~\sqrt{\langle r^2_{\gamma}\rangle |E_3|}= {\cal
R}_{\gamma}^{cm}\left( A\right), \label{rag0}
\end{eqnarray}
depends only on the mass ratio $A$ $(\gamma=n,A)$. Curiously, this
configuration contains simultaneously all types shown in figure 1.

\section{Classification Scheme: qualitative properties}

A discussion of a classification scheme for a bound three-body
system, with two identical particles, is given in ref.~\cite{jensen03},
according to the nature of the subsystem interactions, which can present
a bound or virtual state, as depicted diagrammatically in figure 1.

The different possibilities of three-body systems are reflected in
the qualitative properties of the dynamics as given by the coupled
equations (\ref{chi1}) and (\ref{chi2}) in terms of the strength
of the attractive kernel of these equations. Let us describe all
these possibilities. The {\it Borromean} case corresponds to
positive signs in front of the square-root of the energies of the
subsystems in both eqs. (\ref{tau1}) and (\ref{tau2}). A 
{\it Tango} three-body system has a negative sign only in front of
$\sqrt{\epsilon_{\alpha\alpha}}$ in eq.~(\ref{tau1}), with positive
sign in front of $\sqrt{\epsilon_{\alpha\beta}}$ in
eq.~(\ref{tau2}). For a {\it Samba} configuration of the three-body
system, a minus sign appears multiplying
$\sqrt{\epsilon_{\alpha\beta}}$ in eq.~(\ref{tau2}), while a plus
sign multiplies $\sqrt{\epsilon_{\alpha\alpha}}$ in
eq.~(\ref{tau1}). The All-Bound system has negative signs in
eqs.(\ref{tau1}) and (\ref{tau2}). As a consequence of  the
differences in eqs.(\ref{tau1}) and (\ref{tau2}), the weakest
attractive kernel in eqs.(\ref{chi1}) and (\ref{chi2}) is given by
a {\it Borromean} three-body system, followed by the {\it Tango},
{\it Samba} and {\it All-Bound} systems. Therefore, once fixed a
three-body binding energy, the system for which the kernel
presents the weakest attraction will have the smallest
configuration. So, the  size of the corresponding system will
increase in the following order: {\it Borromean}, {\it Tango},
{\it Samba} and {\it All-Bound}. The {\it All-Bound} configuration
has the biggest size among all, for a given three-body binding
energy. One of course has to remind that the sizes are expected to
increase when the binding energy hits one scattering threshold.
Therefore, for nonvanishing three-body binding energy this
situation does not happen only in the {\it Borromean} case.  In
this sense, it is physically sensible that the {\it Borromean}
case corresponds to the smallest configuration size. This
observation will be explored in our numerical
calculations. In this respect, we are showing the quantitative
detailed implication to the different radii of weakly-bound
three-body systems of the classification scheme proposed in
ref.~\cite{jensen03}.

\section{Numerical Results for the radii: scaling plots}

We present numerical results for the different possible radii for
{\it Borromean}, {\it Tango}, {\it Samba} and {\it All-Bound}
three-body configurations obtained from the wave-function,
eq.~(\ref{psi}) derived from the solution of eqs.~(\ref{chi1}) and
(\ref{chi2}). In particular, we give results for the neutron-neutron
($nn$) average distance in the three-body
halo nuclei, $^6$He, $^{11}$Li, $^{14}$Be, $^{20}$C.

It is interesting to present results in a general scaled form in
terms of scaled two-body binding energies and mass ratio, as given
by eqs.~(\ref{rag}) and (\ref{ragcm}), which turns to be useful
for the prediction of the several radii in different situations
for weakly bound molecules up to light halo-nuclei. Our
calculations present results independent on the detailed form of
the interactions in  a weakly bound three-body system. It means
that they apply very well to halo nuclei and weakly bound
molecules. The present approach is valid as long as  the
interactions within the three-body system are of short range while
the two-body energies are close to zero, i.e., the ratio between
the interaction range and the modulus of the scattering lengths
should be somewhat less than 1. These are indeed the cases we are
considering. We present results only for the ground state of the
three-body system. We have shown that the scaling function for the
radii is indeed approached even for a calculation of the ground
state in our subtraction scheme~\cite{mol03}, which will be
enough for our purpose.

\begin{figure}[thbp]
\centerline{\epsfig{figure=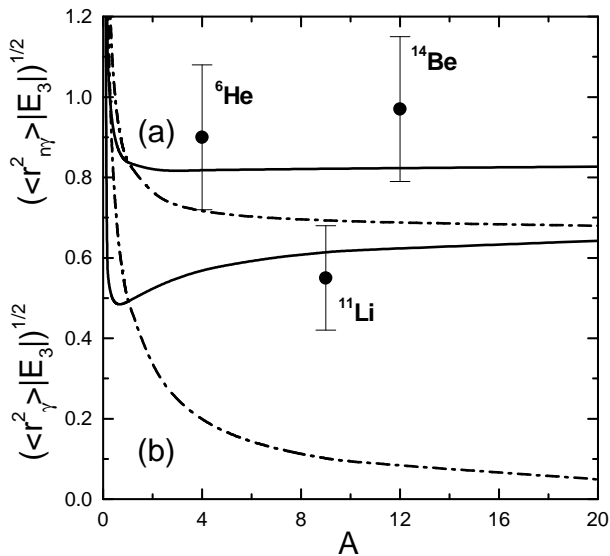,width=8cm}}
\caption[dummy0]{Dimensionless products $\sqrt{\langle
r^2_{n\gamma} \rangle |E_3|}$ and $\sqrt{\langle r^2_{\gamma}
\rangle |E_3|}$ $(\gamma=n,A)$ as a function of $A$ for zero
two-body energies. Results for: (a) $\sqrt{\langle r^2_{nn}
\rangle |E_3|}$ (upper solid line) and $\sqrt{\langle r^2_{nA}
\rangle |E_3|}$ (upper dot-dashed line); (b) $\sqrt{\langle r^2_{n}
\rangle |E_3|}$ (lower solid curve) and $\sqrt{\langle r^2_{A}
\rangle |E_3|}$ (lower dot-dashed curve).
The experimental data, obtained from \cite{marq00}, are for
$\sqrt{\langle r^2_{nn}\rangle |E_3|}$.
}
\label{fig2}
\end{figure}

In figure 2, we show the results of the scaling function for the
mean square distances, $\sqrt{\langle r^2_{n\gamma}\rangle}$ and
$\sqrt{\langle r^2_{\gamma}\rangle}$ with $\gamma=~n$ or $A$ (see
eq.~(\ref{rag0})) which are shown as a function of $A$ for
$E_{n\gamma}=0$.
The comparison with experimental results of $\sqrt{\langle
r^2_{nn}\rangle}$~\cite{marq00}, for the $^6$He, $^{14}$Be and
$^{11}$Li is just for illustrative purpose, considering the
hypothetical cases that $^5$He, $^{13}$Be and $^{10}$Li would have
virtual states close to zero energy. Such hypothesis is more
realistic in case of $^{14}$Be, while for the $^6$He and $^{11}$Li
are not so. 
As shown by our results for $^{11}$Li, given in Table I,
the assumption of a virtual state with energy
close to zero ($|E_{nA}|\lesssim$0.05~MeV~\cite{zin95}) for  
$^{10}$Li lead to an 
average $nn$ separation distance (in the halo of $^{11}$Li)
not compatible with the corresponding experimental result. Given that 
it is well documented the difficulty in studying the $^{10}$Li~\cite{caggi99}, 
we can consider $\sqrt{\langle r^2_{nn}\rangle}$~\cite{marq00}
as one of our inputs to predict the virtual state of
the $n-^9$Li system. In this case, we conclude that it cannot 
be smaller than $\sim$ 0.1 MeV.

In this case ($E_{n\gamma}=0$) the only physical scale is $E_3$ 
and the scaling plots will depend solely on $A$. The average separation 
distance between the neutrons and the neutron-core tends naturally to a
constant for large $A$, while it diverges for $A$ tending towards
zero. The reason for the unbound increase of the products $\langle
r^2_{nn}\rangle |E_3|$ and $\langle r^2_{nA}\rangle |E_3|$ for
small $A$ is due to the average momentum of the core which tends
to zero ($\sim \sqrt{A |E_3|}$) extending the system to infinity.
We have checked that the results shown in figure 2 tends to finite
values after multiplication by $\sqrt{A}$ (one has to remember
that we are using units of $m_n=1$).

The average distance of the neutron to the center of mass system
tends to become the relative distance to the core, when the
core mass grows to infinity. This fact  is clearly seen in figure
2 with the lower solid line approaching the upper dot-dashed one
when $A\gtrsim 10$. Also, one sees that the core distance to the
center of mass vanishes with growing $A$ as it should be.

{\bf
\begin{table}[htb]
\caption[dummy0]
{Results of the neutron-neutron root-mean-square radii in halo nuclei.
The cores ($A$) are given in the first column, the absolute values of the
three-body ground state energies $E_3$ are given in the second column.
For $-E_3$, which is equal to the two-neutron separation energy $S(2n)$,
we consider the center value of the corresponding energies given in
ref.~\cite{audi95}, except for Lithium. In case of Lithium
we consider for $-E_3$ the maximum (0.32 MeV) and the center value 
(0.29 MeV)
given in ref.~\cite{tan96}. In the third column we give our input
values for $-E_{nA}$, considering several values, covering the values
suggested in the literature (the references are given together with
the corresponding number). For bound two-body subsystem $nA$, we have
$-E_{nA}$ equal to the one-neutron separation energy $S(n)$.
The virtual states are indicated by (v), and the $nn$ virtual state
energy is taken as -143 keV. The experimental values, in the last
column, are obtained from Marqu\'es et al. (2000)\cite{marq00}.}\vskip 0.5cm
\begin{tabular}{|c|c|cc|c|c|}
\hline\hline
Core & $-E_3$ & $-E_{nA}$ && $\sqrt{\langle r_{nn}^2\rangle}$
& $\sqrt{\langle r_{nn}^{2}\rangle}_{exp}$ \\(A) &(MeV)&(MeV)&&(fm)&(fm) \\
\hline \hline
      &     &0                &(v)&5.1&\\
$^4$He&0.973&0.3              &(v)&4.6&5.9$\pm$1.2\\
      &     &4.0~\cite{aj88}  &(v)&3.6&\\
\hline
$^9$Li&0.32 &0                     &(v)& 9.2 &6.6$\pm$1.5\\
      &     &0.8 \cite{wil75}      &(v)& 5.9 &           \\
\hline
      &     &0                     &(v)& 9.7 &\\
$^9$Li&0.29 &0.05 \cite{zin95,bar01,tho99}&(v)& 8.5 &6.6$\pm$1.5\\
      &     &0.8 \cite{wil75}      &(v)& 6.7 &           \\
\hline
$^{12}$Be&1.337&0              &(v)&4.6&5.4$\pm$1.0\\
         &     &0.2\cite{tho00}&(v)&4.2&           \\
\hline
$^{18}$C& 3.50&0.16\cite{audi95}&& 3.0 & - \\
         &     &0.53\cite{naka99}&& 4.4 & - \\
\hline\hline
\end{tabular}
\end{table}
}

\begin{figure}[thbp]
\centerline{\epsfig{figure=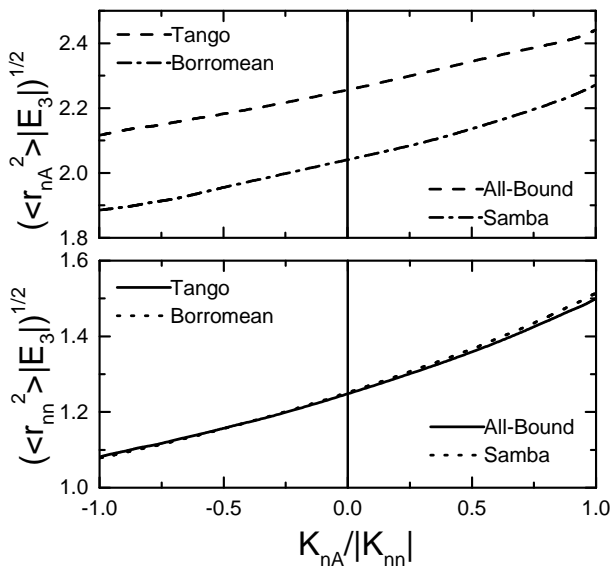,width=8cm}}
\caption[dummy0]{
Dimensionless products $\sqrt{\langle r^2_{nA} \rangle |E_3|}$
(upper frame) and $\sqrt{\langle r^2_{nn} \rangle |E_3|}$ (lower
frame) for $A=0.1$ and $E_{nn}/E_3=K_{nn}^2=0.1$ as a function of
${K_{nA}/|K_{nn}|}$.
In the upper frame, a bound $nn$ pair is represented with dashed line;
and a virtual $nn$ pair with dot-dashed line.
In the lower frame, a bound $nn$ pair is represented with solid line;
and a virtual $nn$ pair with dotted line.
The transition between the configurations occurs when $K_{nA}=0$
(represented by the vertical line).
} \label{fig3}
\end{figure}

\begin{figure}[thbp]
 \centerline{\epsfig{figure=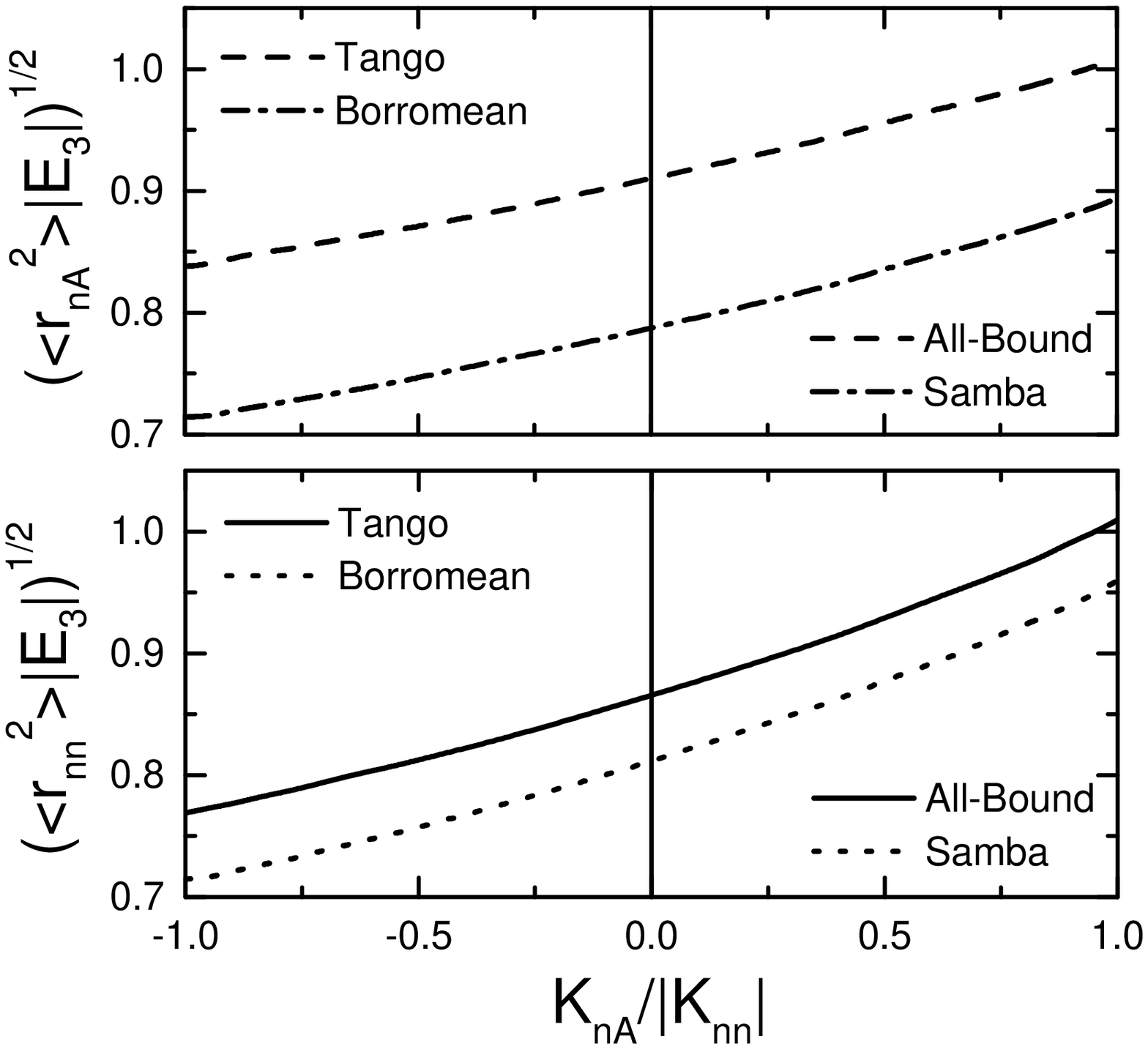,width=8cm}}
\caption[dummy]{
Dimensionless products $\sqrt{\langle r^2_{nA}
\rangle |E_3|}$ (upper frame) and $\sqrt{\langle r^2_{nn} \rangle
|E_3|}$ (lower frame) for $A=1$ and $K_{nn}^2=0.1$ as a function of
${K_{nA}/|K_{nn}|}$.
The convention of the lines is the same as given in figure 3.
} \label{fig4}
\end{figure}

\begin{figure}[thbp]
\centerline{\epsfig{figure=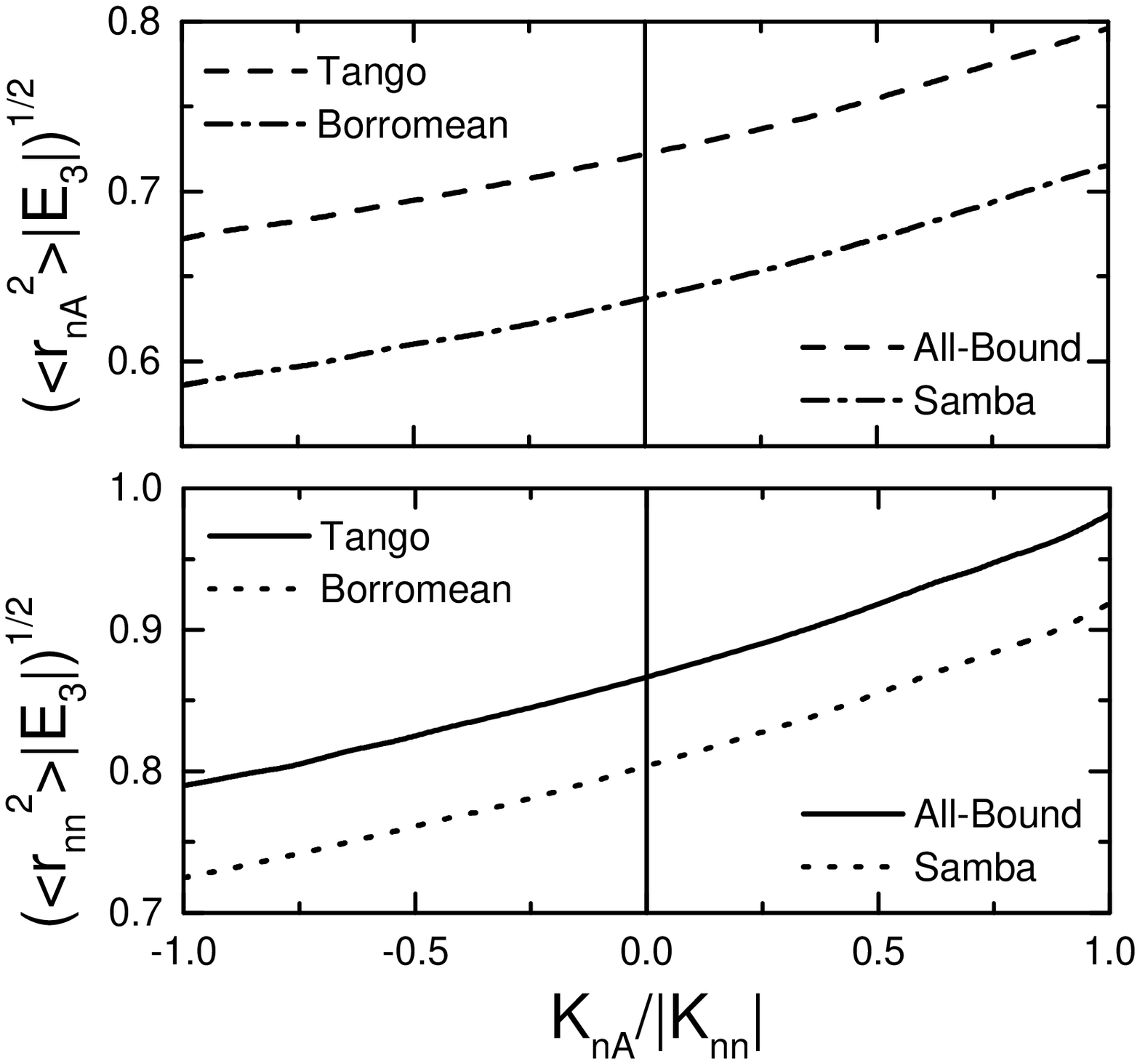,width=8cm}}
\caption[dummy0]{
Dimensionless products $\sqrt{\langle r^2_{nA} \rangle |E_3|}$
(upper frame) and $\sqrt{\langle r^2_{nn} \rangle |E_3|}$ (lower
frame) for $A=200$ and $E_{nn}/E_3=0.1$ as a function of
${K_{nA}/|K_{nn}|}$. The convention of the lines is the same as given
in figures 3 and 4.
} \label{fig5}
\end{figure}

\begin{figure}[thbp]
\centerline{\epsfig{figure=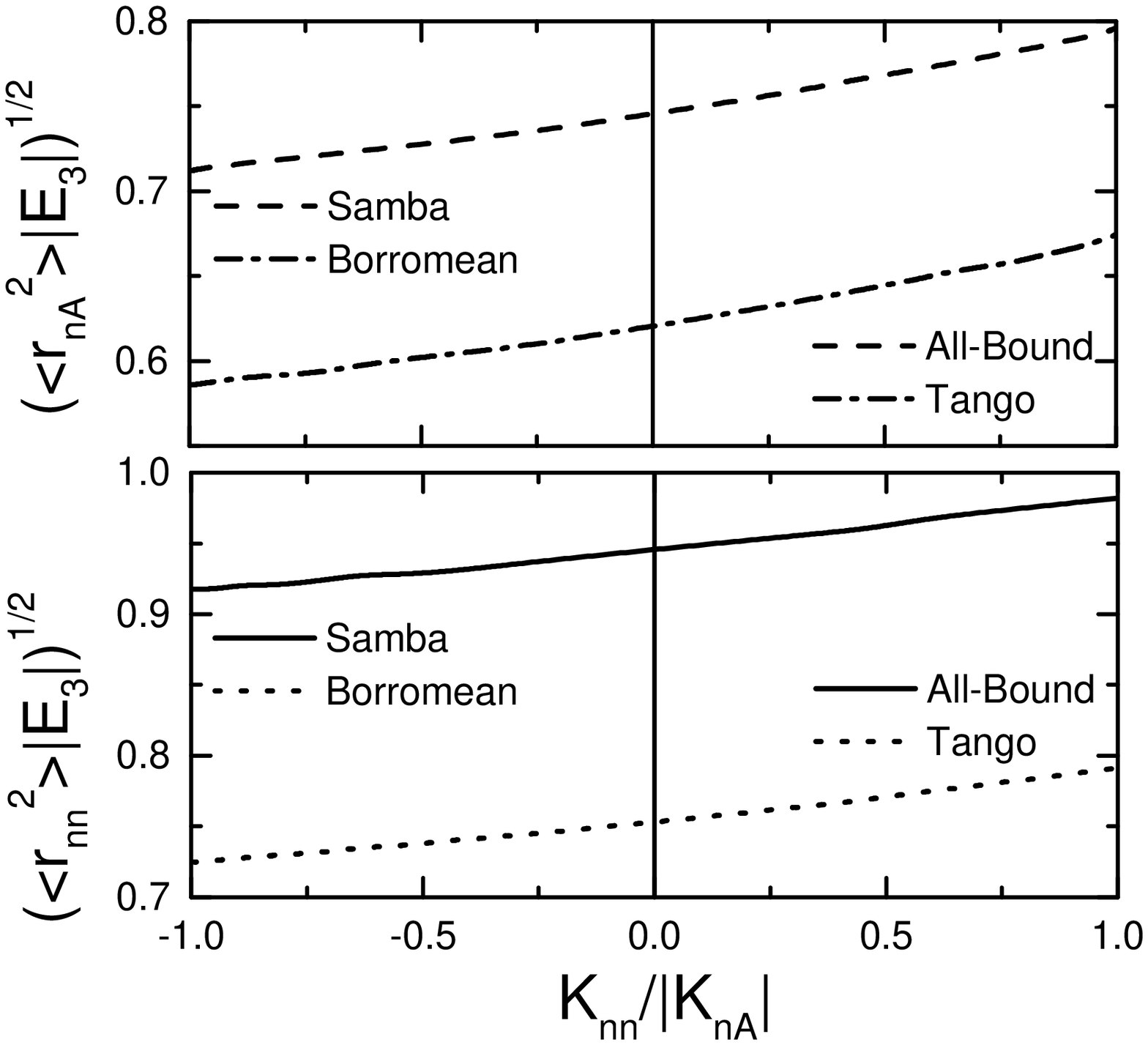,width=8cm}}
\caption[dummy0]{
Dimensionless products $\sqrt{\langle r^2_{nA} \rangle |E_3|}$
(upper frame) and $\sqrt{\langle r^2_{nn} \rangle |E_3|}$ (lower
frame) for $A=200$ and $E_{nA}/E_3=0.1$  as a function of
${K_{nn}/|K_{nA}|}$. In the upper we represent the case of a
bound $nA$ pair with dashed line; and the case of a
virtual $nA$ pair with dot-dashed line. In the lower frame we
represent the bound $nA$ pair with solid line and the
virtual $nA$ pair with dotted line.
The lines of this figure correspond to the vertical transitions
represented in figure 5 in both extreme sides where
$|K_{nA}|=|K_{nn}|$.
} \label{fig6}
\end{figure}
In the figures 3 to 6, we show results for $ \sqrt{\langle
r^2_{nn} \rangle |E_3|}$ and $ \sqrt{\langle r^2_{nA}\rangle
|E_3|}$, when one of the subsystem energies is fixed in respect to
the three-body binding energy, while the other one varies. We use
values of $A$ equal to 0.1, 1 and 200. The subsystems can be bound
or virtual and therefore all configurations are covered, i.e., we
show results for  {\it Borromean}, {\it Tango}, {\it Samba} and
{\it All-bound}-type systems. Anticipating the presentation of
these figures, in general we find that, the radii increase in the
following  order {\it Borromean}, {\it Tango}, {\it Samba} and
{\it All-bound} for a given three-body energy and fixed $A$. In
our analysis below, we fix either  $E_{nn}/E_3=0.1$ or
$E_{nA}/E_3=0.1$ which can correspond to bound or virtual
subsystem energies. 
In the next, we also consider the definitions
$E_{nA}/E_3\equiv K^2_{nA}$ and $E_{nn}/E_3\equiv K^2_{nn}$ (for
both bound or virtual-state energies), such that
\begin{equation}
K_{nA}= \pm\sqrt{E_{nA}/E_3}\;\;\;\;{\rm and}\;\;\;
K_{nn}= \pm\sqrt{E_{nn}/E_3}
\label{K}
.\end{equation}
The $+(-)$ sign refers to bound(virtual) state.
One should also note that, the dimensionless quantities $K_{nn}$ and 
$K_{nA}$ are directly related to poles in the imaginary axis 
of the respective two-body momentum space.

In figure 3, we present calculations of the $nA$ and $nn$
root-mean-square radius as functions of $K_{nA}/|K_{nn}|$. Such
average radius, multiplied by $\sqrt{|E_3|}$, are scaled to
dimensionless quantities. We consider $A=0.1$ and $E_{nn}/E_3=0.1$
fixed, corresponding to bound ($K_{nn}=\sqrt{0.1}$) or virtual
($K_{nn}=-\sqrt{0.1}$) subsystems. In this case, the mass of the
particle $n$ is much heavier than the ``core", which does  not
happen in halo nuclei. However, we consider this case for the sake
of general interest.
In the $x-$axis, the positive (negative) values of $K_{nA}/|K_{nn}|$
correspond to bound (virtual) $nA$ states. In the upper frame,
the average $nA$ radius is shown for a bound $nn$ pair (dashed
line) and for a virtual $nn$ pair (dot-dashed line).
So, the dashed line shows that the value of
$\sqrt{\langle r^2_{nA}\rangle |E_3|}$ increases with the transition
from {\it Tango} to {\it All-bound} configuration.
In the other possibility, represented by the dot-dashed line
($nn$ in a virtual state),
the value of $\sqrt{\langle r^2_{nA}\rangle|E_3|}$ increases from
the most compact {\it Borromean} configuration ($K_{nA}$ negative)
to the {\it Samba}-type configuration ($K_{nA}$ positive).

In the lower frame of figure 3, we also show that
$\sqrt{\langle r^2_{nn}\rangle |E_3|}$ increases with the transition
from {\it Tango} to {\it All-bound} and {\it Borromean} to {\it Samba}
configurations. We observe, in this case, a small sensitivity on
$K_{nn}$, when going from $-\sqrt{0.1}$ (dotted line) to
$\sqrt{0.1}$ (solid line), with
$\sqrt{\langle r^2_{nn}\rangle |E_3|}$ having practically the same
value for the All-bound and Samba configurations; and also for the
Tango and Borromean configurations. As we increase $A$, this
sensitivity increases, as one can see in the next figures 4 and 5.
We interpret this as the dominant role played by a light
particle in the long-range interaction between two heavy
particles, $n$, as already shown in an adiabatical approximation
of the three-body system~\cite{fonseca79}.

In figure 4, we show the radii for the ground state of the
three-body system with the mass ratio $A=1$. As in figure 3, we
fixed $K_{nn}=\pm\sqrt{0.1}$, corresponding to bound ($+$) or virtual 
($-$) subsystems. The same behavior found in figure 3 for 
$\sqrt{\langle r^2_{nA}\rangle |E_3|}$ is found in the upper frame of 
figure 4 for $A=1$, i.e., the
configurations for which the $nn$ pair is virtual (dot-dashed
line) are smaller than the ones that have the $nn$ pair bound
(dashed line). Besides that, the configuration increases in size
when the system changes from a {\it Borromean} to a {\it Samba}
type and when it changes from a {\it Tango} to an {\it All-bound}
type. For $A=1$, the mean square distance between the two-neutrons
(lower frame of figure 4) exhibits the same qualitative
behavior as found for  $\sqrt{\langle r^2_{nA}\rangle |E_3|}$ when
the configuration type is modified for a fixed three-body energy.
These conclusions are still valid for a heavy core particle with
$A=200$, as one can verify in figure 5. It is worthwhile to
mention that $\sqrt{\langle r^2_{nn}\rangle |E_3|}$ attains its
asymptotic value fast with the increase of $A$ consistently with
the calculations presented in figure 2 for $E_{nn}=E_{nA}=0$.
(Compare the results for $A=1$ and $A=200$ in the lower frames of
figures 4 and 5).

In correspondence with figure 5, for the same mass ratio $A=200$,
in the last set of systematic calculations we consider in figure 6
the energy of the subsystem $nA$ fixed in relation to $E_3$, with
$K_{nA}=\pm\sqrt{0.1}$, corresponding to bound ($+$) or
virtual ($-$) $nA$ subsystem.  In this case, the ratio
$K_{nn}/|K_{nA}|$ is changed from $-1$ to $+1$, such that the
transitions of configurations from the left side to the right side
of this figure correspond to the vertical transitions in the
extreme side of figure 5 (when $|K_{nA}|=|K_{nn}|=\sqrt{0.1}$).
So, comparing the upper frames of figures 5 and 6, we observe that
the vertical transition from {\it All-bound} to {\it Samba} configuration
in figure 5 corresponds to the dashed line of figure 6; and,
the vertical transition from {\it Tango} to {\it Borromean} configuration
in figure 5 corresponds to the dot-dashed line of figure 6.
In the lower frames of both figures, similarly we observe that the
vertical transitions of figure 5 correspond to the lines represented in
figure 6: the transition from {\it Samba} to {\it All-bound} configuration
is given by the solid line; and the transition from {\it Borromean} to
{\it Tango} configuration given by the dotted line.
In general, one can observe that the three-body bound state has a
smaller size when the $nA$ pair forms a virtual state (see in each frame
of figure 6, where the upper line is for bound and the lower line is for
virtual $nA$ pair).

The calculations of the average distances of the neutrons in the
halo of $^6$He, $^{11}$Li, $^{14}$Be and $^{20}$C are shown in
Table I and compared with the available experimental data.  For
the input, we have considered $E_{nn}=-0.143$ MeV and the center
of the available experimental values of $E_3$ and $E_{nA}$. For
the cases that we have unbound virtual $nA$ systems, the $E_{nA}$
input values are taken from several recent theoretical and
experimental analysis; we prefer to keep at least two possible
values in order to verify the consistency of the model with the
experimental data.

Within the possible limitations of our approach,
by comparing our results for the $nn$ mean-square radius with the
experimental ones, which are known in the cases of
$^6$He, $^{11}$Li and $^{14}$Be, as given in Table 1,
one can also predict the corresponding virtual energies of the
$nA$ subsystem.

In the particular case of $^6$He, the comparison between the
experimental $nn$ mean-square radius with our result is pointing
out to a virtual energy close to zero for $n-^4$He, which is not
supported by the quite large values given in the
literature~\cite{zheng93,aj88}. However, as discussed in
\cite{zheng93,aj88}, the interaction for $n-^4$He is
attractive in $p-$wave and repulsive in $s-$wave producing a large
value for the energy of the virtual state, such that a deviation
of our calculation from the experimental result is expected (in
our model the  $s-$wave poles should be near the scattering
thresholds). For the binding of $^{11}$Li, it is also known that
both $p-$ and $s-$waves are important in the subsystem $n-^9$Li.
This fact can also explain some deviations of our results when
compared with experimental ones. However, in this last case,  our
approach can be more reliable based on the fact that: {\it(i)}
the $s-$wave $n-^9$Li interaction is attractive and it has a
virtual state near the scattering region; {\it(ii)} we are
considering the experimental energies for the inputs, such that we
are partially taking into account the effect of higher partial
waves in the interactions; {\it(iii)} the radius is obtained from
the tail of the wave-function, which is dominated by the $s-$wave;
{\it(iv)} as noted in ref.~\cite{zin95}, even the three-body wave
function with $s-$wave $nn$ correlation produces a ground state
of the halo nuclei with two or more shell-model configurations.

One should also expect that other effects like the finite
size of the core and Pauli principle, missing in our model, could
affect the average relative distances, if the three-body wave
function would overlap appreciably with the core. At least
for $^{11}$Li this is not expected due to the small binding.
It is of notice that indeed the halo neutrons have a large
probability to be in a region  in which the wave function is an
eigenstate of the free Hamiltonian, and thus dominated by few
asymptotic scales.

\section{Conclusions}

The mean-square radii of a light halo-nuclei modelled as a
three-body system (two neutrons $n$ and a core $A$) are calculated
using a renormalized three-body model with a pairwise
Dirac-$\delta$ interaction, which works with a minimal number of
physical inputs directly related to observables. These physical
scales are the two-neutron separation energy $S(2n)=-E_3$, and the
$nn$ and $n-$core $s-$wave scattering lengths.

The existent data for $^{11}$Li and $^{14}$Be compare qualitatively well
with our theoretical results, which means that the neutrons of the halo
have a large probability to be found outside the interaction
range. Therefore the low-energy properties of these halo
neutrons are, to a large extend, model independent as long as few
physical input scales are fixed.  
The model provides a good insight into the three-body structure 
of halo nuclei, even considering some of its limitations.
We pointed out that the model is restricted to $s-$wave subsystems,
with small energies for the bound or virtual states.
So, the model is not suitable for the $^6$He, since the $s-$wave 
virtual state energy of $^5$He is quite large (-4~MeV).
There is no $n-$core $p-$wave interaction, although some
of its physics is effectively included through the value of the
two-neutron separation energy, which is an input for our radii
calculations. Also the finite size of the core and consequently
the antisymmetrization of the total nuclear wave function, are
both missing in our model. However, as the three-body halo nuclei
tend to be more and more weakly bound with subsystems that have
bound or virtual states near the scattering threshold, our
approach becomes adequated and the above limitations are softened.
The results indicate that the model is reasonable for 
$^{11}$Li and $^{14}$Be.

As an example of application to other halo-nuclei system, 
considering the available energies, we have also estimated the 
$nn$ root-mean-square radius for the $n-n-^{18}$C system.

We have also studied in detail the consequences of the
classification scheme proposed in Ref.~\cite{jensen03} for weakly
bound three-body systems. This study was performed analyzing the
dimensionless products $\sqrt{\langle r^2_{nA} \rangle |E_3|}$ and
$\sqrt{\langle r^2_{nn} \rangle |E_3|}$  in terms of scaling
functions depending on the dimensionless product of  the
scattering lengths and the square-root of the neutron-neutron
separation energy. In the cases we have addressed, there are four
different types of a three-body system when we allow the
neutron-neutron pair to be bound: {\it Borromean} (only virtual
two-body subsystems), {\it Tango} ($nn$ bound and $nA$ virtual),
{\it Samba} ($nn$ virtual and $nA$ bound) and {\it All-Bound}
(only bound two-body subsystems). The name {\it Samba} was
introduced  to refer to  a system quite stable because it has two
 bound two-body subsystem than the {\it Tango} type, so it can
``shake" a little bit more and continue to be bound.

The qualitative properties of the different possibilities of
three-body systems are easily understood in terms of the effective
attraction in the model: when a pair has a virtual state the
effective interaction is weaker than when the pair is bound. Thus,
a three-body system has to shrink to keep the binding energy
unchanged if a pair which is bound turns to be virtual. We have
illustrated through several examples this property, which show
that dimensionless sizes $\sqrt{\langle r^2_{nA} \rangle |E_3|}$
and $\sqrt{\langle r^2_{nn} \rangle |E_3|}$ increase from {\it
Borromean}, {\it Tango}, {\it Samba} and to {\it All-Bound}
configurations. Of course the size is expected to increase beyond
limits when a nonvanishing three-body energy hits a scattering
threshold, with the {\it Borromean} configuration being the only
exception. In spite of that, we conclude that even far from the
threshold situation, the configuration sizes increase as we
pointed out.

We would like to thank Funda\c c\~ao de Amparo \`a Pesquisa do
Estado de S\~ao Paulo (FAPESP) for partial support. LT and TF also
thank partial support from Conselho Nacional de Desenvolvimento
Cient\'{\i}fico e Tecnol\'ogico (CNPq).


\begin{thebibliography}{}
\bibitem{zhukov93}
M.V. Zhukov, B.V. Danilin, D.V. Fedorov, J.M. Bang,
I.J. Thompson, J.S. Vaagen, Phys. Rep. 231 (1993) 151.

\bibitem{nielsen01} E. Nielsen, D.V. Fedorov, A.S. Jensen and
E. Garrido, Phys. Rep. 347 (2001) 373.

\bibitem{audi95}
G. Audi and A.H. Wapstra, Nucl. Phys. A 595 (1995) 409.

\bibitem{amorim97} A.E.A. Amorim, L. Tomio and T. Frederico,
Phys. Rev. C 56 (1997) R2378.

\bibitem{efimov70}
Efimov, V.: Phys. Lett. B 33 (1970) 563.

\bibitem{fedorov94}
D. V. Fedorov, A. S. Jensen, and K. Riisager
Phys. Rev. Lett. 73 (1994) 2817.

\bibitem{mazum00}
I. Mazumdar, V. Arora, and V. S. Bhasin
Phys. Rev. C 61 (2000) 051303.

\bibitem{ad95} S.K. Adhikari, T. Frederico and I.D. Goldman,
Phys. Rev. Lett.  74 (1995) 487; S.K. Adhikari and T.
Frederico, Phys. Rev. Lett. 74 (1995) 4572.

\bibitem{fred99}
T. Frederico, L. Tomio, A. Delfino, and A. E. A. Amorim, Phys.
Rev. A 60 (1999) R9.

\bibitem{ad88}
S. K. Adhikari, A. Delfino, T. Frederico, I. D. Goldman, and
L.Tomio, Phys. Rev. A 37 (1988) 3666.

\bibitem{fb17}
M.T. Yamashita, T. Frederico, R.S. Marques de Carvalho, and
L. Tomio, {\it Radii of weakly bound three-body systems: halo nuclei
and molecules}, nucl-th/0308072,
to appear in Nucl. Phys. A.

\bibitem{jensen03} A.S. Jensen, K. Riisager, D.V. Fedorov and E. Garrido,
Europhys. Lett. 61 (2003) 320.

\bibitem{robi99}
F. Robicheaux, Phys. Rev. A 60 (1999) 1706.

\bibitem{naka99}
T. Nakamura et al., Phys. Rev. Lett. 83 (1999) 1112.

\bibitem{mol03} M.T. Yamashita, R.S. Marques de Carvalho, L.
Tomio and T. Frederico, Phys. Rev. A 68 (2003) 012506.

\bibitem{marq00}
F.M. Marqu\'es et al., Phys. Lett. B 476 (2000) 219;
Phys. Rev. C 64 (2001) 061301.

\bibitem{yama02} M.T. Yamashita, T. Frederico, A. Delfino, and
L. Tomio, Phys. Rev. A 66 (2002) 052702.

\bibitem{zheng93} D.C. Zheng et al., Phys. Rev. C 48 (1993) 1083.

\bibitem{caggi99} J.A. Caggiano et al., Phys. Rev. C 60 (1999) 064322.

\bibitem{zin95} M. Zinser et al., Phys. Rev. Lett. 75 (1995) 1719;
M. Zinser et al., Nucl. Phys. A 619 (1997) 151.

\bibitem{fred87}T. Frederico, I. D. Goldman, Phys. Rev. C 36
(1987) R1661.

\bibitem{tan96} I. Tanihata, J. Phys. G22 (1996) 157.

\bibitem{aj88} F. Ajzenberg-Selove, Nucl. Phys. A490 (1988) 1.

\bibitem{wil75} K.H. Wilcox et al., Phys. Lett. 59B (1975) 142.

\bibitem{bar01} F. Barranco, P.F. Bortignon, R.A. Broglia, G. Colo,
E. Vigezzi, Eur. Phys. J. A 11 (2001) 385.

\bibitem{tho99} M. Thoennessen et al. Phys. Rev. C 59 (1999) 111.

\bibitem{tho00} M. Thoennessen, S. Yokoyama, P.G. Hansen, Phys. Rev. C
63 (2000) 014308. 

\bibitem{fonseca79} A. C. Fonseca, E. F. Redish and P. E. Shanley,
Nucl. Phys. A 320 (1979) 273.

\end{thebibliography}
\end{document}